\documentclass[12pt]{article}
\usepackage[OT1]{fontenc}
\usepackage{verbatim,color,epsfig,setspace}
\usepackage{mathrsfs}
\usepackage{amsmath}
\usepackage{amsthm}
\usepackage{graphicx}
\usepackage{amsfonts}
\usepackage{amssymb}
\usepackage{float}
\usepackage{url}
\makeatletter
\g@addto@macro{\UrlBreaks}{\UrlOrds}
\makeatother
\PassOptionsToPackage{hyphens}{url}\usepackage{hyperref}
\usepackage{epic}
\usepackage[authoryear]{natbib}

\newif\ifunderreview
\underreviewtrue

\newif\ifpriorwg
\priorwgfalse

\newtheorem{Def}{Definition}

\newtheorem{proposition}{\bf {Proposition}}
\newtheorem{lemma}{{\bf Lemma}}

\def\bmath#1{{\boldsymbol{#1}}}

\def\Y{\mathbf{Y}}                          
\def\y{\mathbf{y}}
\def\X{\mathbf{X}}                          

\def\G{\mathbf{G}}                          
\def\SG{\mathbf{\Sigma}_{\G}}               
\def\I#1{\mathbf{I}_{#1}}                   
\def\W{\mathbf{W}}                          
\def\setG{\mathfrak{G}}                     
\def\PG{P_{\G}}                             
\def\uv{uv}                                 
\def\udv{\left\{u,v\right\}}                
\def\V{V}                                   
\def\E{\mathbf{E}}                          

\def\t{\mathbf{t}}
\def\uvinE{\udv\in\E}                        

\def\y{\mathbf{y}}                          
\def\tt{\tilde{\t}}                         

\newcommand{\0}{{\bf 0}}
\renewcommand{\I}{{\bf I}}
\newcommand{\D}{{\bf D}}

\def\MN#1#2#3#4#5{\mathsf{MN}_{#1\times#2}\left(#3,#4,#5\right)}       
\def\HIW#1#2#3{\mathsf{HIW}_{#1}\left(#2,#3\right)}                    
\def\HMT#1#2#3#4#5{\mathsf{HMT}_{#1\times#2}\left(#3,#4,#5\right)}     

\def\cond{\ \vert\ }                        
\def\iid{\mathrm{i.i.d.}}                   
\def\simiid{\stackrel{\iid}{\sim}}          

\def\boxit#1{\vbox{\hrule\hbox{\vrule\kern6pt
          \vbox{\kern6pt#1\kern6pt}\kern6pt\vrule}\hrule}}

\def\log{\hbox{log}}

\def\bse{\begin{eqnarray*}}
\def\ese{\end{eqnarray*}}
\def\be{\begin{eqnarray}}
\def\ee{\end{eqnarray}}
\def\bq{\begin{equation}}
\def\eq{\end{equation}}
\def\bse{\begin{eqnarray*}}
\def\ese{\end{eqnarray*}}

\def\b1e{{\mathbf e}}
\def\bx{{\mathbf x}}
\def\bz{{\mathbf z}}
\def\bX{{\mathbf X}}

\def\D{{\mathbf D}}

\def\bY{{\mathbf Y}}
\def\Z{{\mathbf Z}}

\newcommand{\bSigma}{\boldsymbol{\Sigma}}

\newcommand{\bOmega}{\mbox{\boldmath $\Omega$}}

\def\summ3{\hbox{$\sum_{\ell \neq 3}$}}

\def\bG{\mathbf G}

\def\bSigma{\mathbf\Sigma}
\def\bI{\mathbf I}
\def\bx{\mathbf x}
\def\by{\mathbf y}

\newcommand{\ben}{\begin{enumerate}}
\newcommand{\een}{\end{enumerate}}
\newcommand{\beq}{\begin{equation}}
\newcommand{\eeq}{\end{equation}}

\allowdisplaybreaks

\def\boxit#1{\vbox{\hrule\hbox{\vrule\kern6pt
          \vbox{\kern6pt#1\kern6pt}\kern6pt\vrule}\hrule}}

\newcommand{\Var}{\mathrm {Var}}

\newcommand{\N}{\mathrm{N}}

\newcommand{\bGamma}{\boldsymbol{\Gamma}}
\newcommand{\bTheta}{\boldsymbol{\Theta}}

\def\b1e{{\mathbf e}}
\def\bx{{\mathbf x}}
\def\bX{{\mathbf X}}

\def\D{{\mathbf D}}

\def\bY{{\mathbf Y}}
\def\bZ{{\mathbf Z}}

\def\k {\mathcal \rho}

\begin{document}
\thispagestyle{empty}
{\baselineskip=22pt \LARGE{\bf  \noindent \bf Inferring network structure in non-normal and mixed discrete-continuous genomic data}}
\date{}
\vskip 10mm
\baselineskip=12pt
\begin{center}
Anindya Bhadra\\
Department of Statistics, Purdue University, 250 N. University St., West Lafayette, IN 47907\\
bhadra@purdue.edu\\
\hskip 5mm \\
Arvind Rao\\
Department of Bioinformatics and Computational Biology, The University of Texas MD Anderson Cancer Center, 1400 Pressler Dr., Houston, TX 77030\\
\hskip 5mm \\
Veerabhadran Baladandayuthapani\\
Department of Biostatistics, The University of Texas MD Anderson Cancer Center, 1400 Pressler Dr., Houston, TX 77030  \\
\end{center}
\vspace{5mm}
\begin{center}
{\Large{\bf Abstract}}
\end{center}
{\baselineskip=20pt
\noindent  Inferring dependence structure through undirected graphs is crucial for uncovering the major modes of multivariate interaction among high-dimensional genomic markers that are potentially associated with cancer. Traditionally, conditional independence has been studied using sparse Gaussian graphical models for continuous data and sparse Ising models for discrete data. However, there are two clear situations when these approaches are inadequate. The first occurs when the data are continuous but display non-normal marginal behavior such as heavy tails or skewness, rendering an assumption of normality inappropriate. The second occurs when a part of the data is ordinal or discrete (e.g., presence or absence of a mutation) and the other part is continuous (e.g., expression levels of genes or proteins).  In this case, the existing Bayesian approaches typically employ a latent variable framework for the discrete part that precludes inferring conditional independence among the data that are \emph{actually observed}. The current article overcomes these two challenges in a unified framework using Gaussian scale mixtures. Our framework is able to handle continuous data that are not normal and data that are of mixed continuous and discrete nature, while still being able to infer a sparse conditional sign independence structure among the observed data. Extensive performance comparison in simulations with alternative techniques and an analysis of a real cancer genomics data set demonstrate the effectiveness of the proposed approach.}
\baselineskip=12pt
\baselineskip=12pt
\par\vfill\noindent
\par\medskip\noindent
\clearpage\pagebreak\newpage
\pagenumbering{arabic}
\newlength{\gnat}
\setlength{\gnat}{16pt}
\baselineskip=\gnat
\section{Introduction}
With rapid advances in high-throughput genomic technologies using array and sequencing-based approaches, it is now possible to collect detailed high-resolution molecular information across the entire genomic landscape at various levels. The data can be genetic (e.g, mutations or single neucleotide polymophisms), genomic (e.g., expression levels of messenger RNA and microRNA), epigenomic (e.g., DNA methylation) or proteomic (e.g., protein expression). 
The interrelations among these data provide key insights into the etiology of many diseases, including cancer. 
Statistically, the question of uncovering the major modes of multivariate interactions 
in genomic data can be phrased in terms of  inferring a \emph{conditional independence} graph. A unifying feature of these genomics problems is that the number of variables ($q$) far exceeds the sample size ($n$). Therefore, a multivariate sparse Gaussian graphical model is commonly applied to analyze the conditional independence structure \citep[see, e.g.,][]{lauritzen96, carvalho07, friedman08, meinhausen06}. Given this high-dimensional setting, the purpose of the current article is to study multivariate interactions in two important situations where a Gaussian graphical model is inappropriate. These are (i) when the data are continuous, but display non-normal features such as heavy tails or skewness and (ii) when  the data are of mixed discrete and continuous nature.

\begin{figure}[!t]
\begin{center}
\includegraphics[height=8cm,width=16cm]{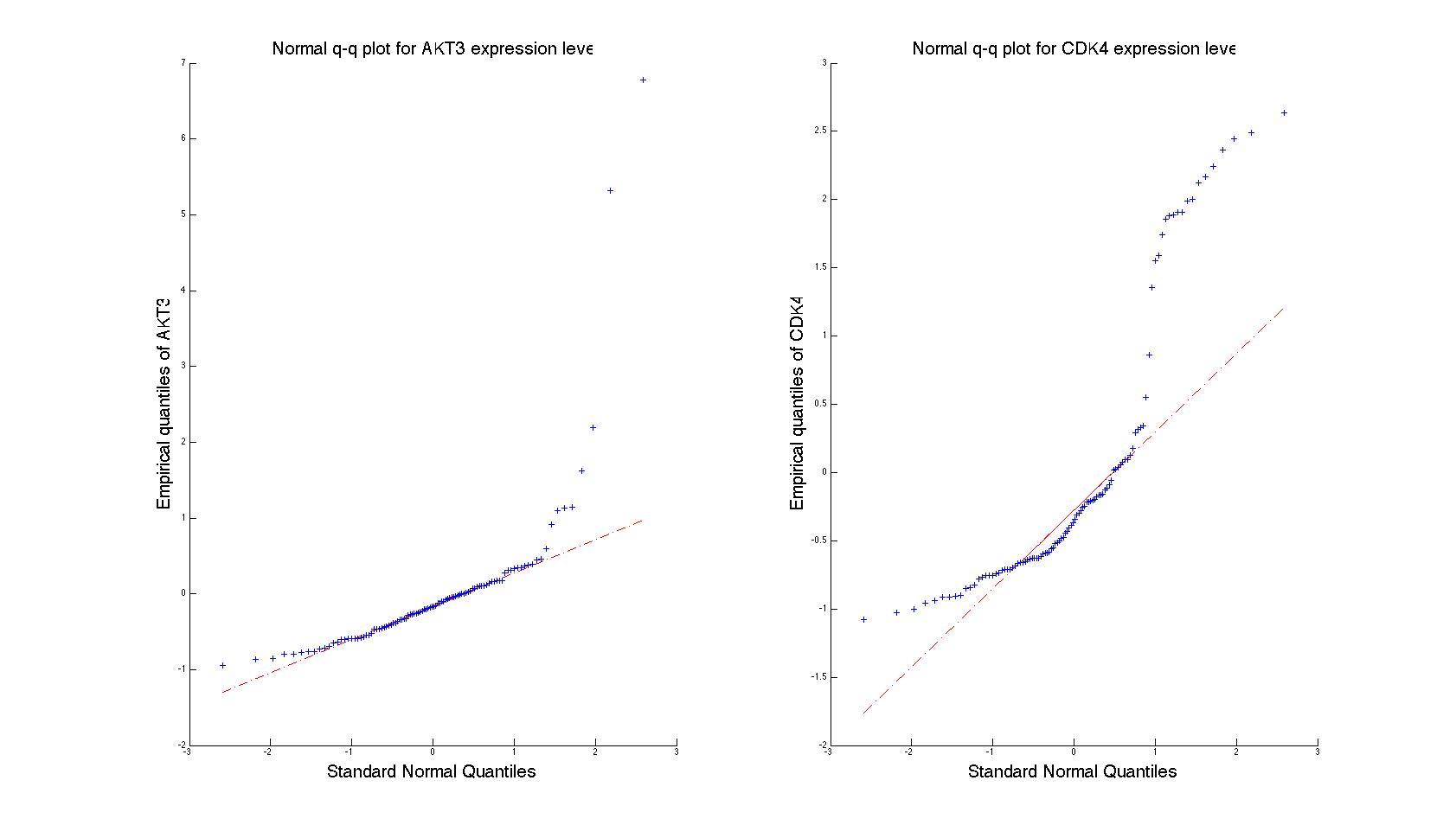}
\end{center}
\caption{An illustration of non-normal marginals in genomic data. Normal q-q plots for the marginals of AKT3 and CDK4 expression levels based on TCGA glioblastoma samples, clearly demonstrating non-Gaussian tails. These two genes have been implicated in glioblastoma by \citet{tcga}. \label{fig:tcga}}
\end{figure}

First, consider the case where all data are continuous but possibly non-normal. This is particularly important in genomics where the data often display features such as heavy tails. Moreover, in a multivariate setting, each marginal may display a separate characteristic.  As a motivating example, in Figure~\ref{fig:tcga} we plot the expression levels of two genes (AKT3 and CDK4) that are implicated in glioblastoma multiforme (GBM), which is the most aggressive form of brain cancer \citep{tcga}. It is apparent that each marginal deviates from normaility \emph{in a different way}, especially in the tails (Kolmogorov-Smirnov test p-values 6.26e-6 and 1.49e-4, respectively). Since diseases such as cancer are often characterized by extreme changes in gene expression \citep{gray01}, capturing the tail behavior is crucial. Biological consequences of using a misspecified Gaussian model are serious, potentially resulting in an inference of wrong associations  \citep{marko12}. There are some recent works in Bayesian literature for allowing for more flexible marginal behavior in the data, e.g., the alternative multivariate-$t$ or Dirichlet-$t$ of \citet{finegold11, finegold14}, but, in view of Figure~\ref{fig:tcga}, it raises the question why one particular distribution (e.g., a $t$-distribution) would be appropriate along all the marginals. Furthermore, a $t$-distributed marginal cannot model important behavior often observed in genomics, e.g., skewness.

A second problem with genomic data is that it is heterogeneous (mixed discrete, ordinal and continuous).  For example, presence or absence of mutations are modeled as binary variables; copy number aberrations as ordinal variables (gain/loss/normal); and expression levels of microRNA or messenger RNA are continuous. Characterizing the dependence among heterogeneous types of data is not well-understood, even in low dimensions. A typical Bayesian approach is to model the discrete part with  latent continuous random variables and then to infer the conditional independence structure among the observed and latent continuous variables. It is unclear, however, how this latent dependence or correlation translates to the observed data \citep{pitt06}. Outside of Bayesian approaches, this problem has received some recent attention, but the proposed techniques are limited to exponential family of distributions \citep{cheng13, yang15, lee2015learning}.

Given these two problems, the focus of the current work is to delineate a unifying framework that can infer ``conditional sign independence'' in the face of data that are non-Gaussian and are of mixed discrete/continuous nature. We define two random variables $\zeta_1$ and $\zeta_2$ to be conditionally sign independent given $\zeta_3$, if the sign of $\zeta_1$ given $\zeta_3$ remains independent of whether $\zeta_2$ is also known. A more precise definition is given later in Definition~\ref{def:indep}.  Note that this definition has an intuitive appeal in multivariate genomic data of mixed nature. Here it might not make sense to compare the numeric values of data that are truly quantitative (e.g., gene expression) versus data that are binary $\{1, -1\}$ coded dummy variables (presence or absence of a mutation). But one might still be interested to see if positive values of the dummy variable (indicating presence of mutation) co-occurs with positive expression level of some gene (also known as up-regulation), conditional on the rest of the variables of interest. One might also want to investigate if two arbitrarily coded binary  deleterious mutations are likely to co-occur, accounting for the effect of the rest of the variables.

Using a Gaussian scale mixture representation of the marginals, we show that it is possible to draw these conclusions. A key contribution of our work is that we can make statements concerning conditional sign independence among \emph{observed} discrete and continuous random variables. This property makes our approach distinct from the literature on Bayesian copula graphical models \citep[e.g., ][]{pitt06} that can only make statements conditional on some latent variables. The rest of the manuscript is organized as follows. In Section~\ref{sec:bayesg}, we provide the necessary background on Bayesian approaches to Gaussian graphical models. We discuss the two main innovations of the paper, characterization of conditional sign independence in non-Gaussian and mixed discrete-continuous data in Sections~\ref{sec:heavy} and~\ref{sec:mixed} respectively. Simulation results and extensive performance comparison with alternative approaches are in Section~\ref{sec:sim}. We analyze a cancer genomics data set in Section~\ref{sec:real}. We conclude by pointing out some directions of future investigation, including a possible E-M scheme that can be useful in non-Bayesian analysis of mixed data, in Section~\ref{sec:conc}.

\section{Bayesian approaches to Gaussian graphical models} \label{sec:bayesg}
Consider a Gaussian graphical model for purely continuous data of the following form:
\begin{equation}
\Y \sim \MN{n}{q}{\mathbf{0}_{n\times q}}{\I_n}{\SG}
\end{equation}
where $\Y$ is an $n \times q$ data matrix, modeled as a matrix-variate normal \citep{dawid81}. Here $\mathbf{0}_{n\times q}$ is an $n \times q$ mean matrix of zeros, $\SG$ is the $q \times q$ column covariance matrix of $q$ possibly correlated variables and $\I_n$ is an identity matrix of size $n$. The matrix normal formulation implies a separable covariance structure of $\Y$ along the rows and columns and $\mathrm{Vec} (\Y) \sim \mathrm{N}_{nq}({\mathbf{0}_{nq}}, {\I_n}\otimes{\SG})$, a multivariate normal, with $\otimes$ denoting the Kronecker product. This formulation is justified when the $n$ samples are independent, but within each sample, the $q$ responses share a common covariance structure encoded by $\SG$ due to interaction among the variables (e.g., gene interaction network when the variables are gene expressions). Conditional independence is modeled through an underlying (undirected) graph $\G=\left(\V,\E\right)$, where $\V$ corresponds to response variables $Y_1,\dots,Y_q$, with the implication that $\{u,v\} \not\in\E \iff \SG^{-1}(u,v)=0$, implying conditional independence of $u$ and $v$ given the rest, where
$u,v\in\V$. Clearly, when  $q$ is much larger than $n$, the model is not identifiable. Thus, we consider the following hierarchical sparse Bayesian model: 
\begin{eqnarray}
 G_{\uv} &\simiid& \PG\left(\cdot\cond\W\right), \label{eqn:G}\label{eq:gamma}\\
\SG \cond \G &\sim& \HIW{\G}{b}{\k\I_{q}}, \label{eq:sigmag}\\
\Y \cond \SG &\sim&  \MN{n}{q}{\0}{\I_n}{\SG}.\label{eqn:model}
\end{eqnarray}
In Equation \eqref{eqn:G}, we restrict the set of permitted graphs to $\setG$, the set of all decomposable (or, triangulated) graphs with nodes $\V$, and define a distribution with support over $\setG$ as
\begin{equation}
\PG\left(\G\cond \W\right) \propto  \left[ \prod_{\uvinE}w_{uv}\right]\left[\prod_{\udv\not\in\E}\left(1-w_{uv}\right)\right].\label{eqn:PG}
\end{equation}
The model  specifies that the prior on $\SG$ is conjugate in a graphical setting, which allows analytic marginalization. The hyper-inverse Wishart {\rm (HIW)} distribution is a conjugate prior for the covariance matrix in a decomposable Gaussian graphical model \citep{dawid93}. Here $b, \k$ are fixed, positive hyper-parameters.  A symmetric matrix $\W=\left(w_{\uv}\right)_{u,v\in\V}$ are fixed prior weights that control the sparsity in $\G$. For inference on $\G$, one may work with the marginal model with $\SG$ integrated out, which gives
\begin{align*}
\Y \cond  \G  \sim \HMT{n}{q}{b}{\I_n}{\k\I_q}.
\end{align*}
If the graphs $\G\in\setG$ are {\em decomposable}, the distribution of $\Y \cond  \bG$ is {\em hyper-matrix} $t$ \citep[abbreviated as HMT, ][]{dawid93}, a special type of $t$-distribution which, given the graph, splits into products and ratios over the cliques and separators of the graph.
We recall that a decomposable graph $\G$ admits a (perfect) sequence of maximal cliques $C_1,\dots,C_l$ and $S_j = \left(C_1\cup \dots \cup C_{j-1}\right) \cap C_j$, $j=2,\dots,l$ (called {\em separators}) are complete sub-graphs of $\G$ \citep{lauritzen96}.  The density of the hyper-matrix-$t$ distribution $\HMT{n}{q}{b}{\I_n}{\k\I_q}$ is 
\begin{align}
f\left(\y \cond  \G\right)  &= \frac{\prod_{j=1}^{l}f(\y_{C_j}\cond  \G)}{\prod_{j=2}^{l}f(\y_{S_j}\cond  \G)}, \quad\mbox{where}\quad  f\left(\y_{C_j}\cond \G\right) \propto \det\left(\I_{|C_j|} + \y_{C_j}^t\y_{C_j}/\k\right)^{-(b+n+\left|C_j\right|-1)/2},\label{eqn:lik}
\end{align}
 at $\Y=\y$ and $\t_{A}$ is a $n \times \left|A\right|$ sub-matrix of $\t$ with columns corresponding to cliques $A\subseteq \V$ in $\G$ \citep[Equation (45) of][]{dawid93}. Infrence on $\G$ typically proceeds by random addition or deletion of edges in the graph and by computing the appropriate M-H ratio \citep{giudici99, scott08, Bhadra:Mallick:2013, mohammadi2015bayesian}. Additionally, if the posterior estimate of $\bSigma_\bG$ is also desired, one can sample from the conditional distribution as:
\begin{align*}
\SG \cond \Y, \G &\sim \HIW{\G}{b + n}{\k\I_{q} + \bY'\bY}.
\end{align*}
We note that in order to infer conditional independence it is actually not necessary to restrict oneself to decomposable graphs. One can work with more general G-inverse Wishart priors on $\bSigma_\bG$ instead of HIW. Samplers for non-decomposable graphs \citep{wang2010} or mixtures of tree-structured graphs \citep{feldman2014bayesian} exist, although they are not as computationally efficient and do not always scale well to high dimensions. Thus, we use the framework of decomposable models, although it is not strictly required for the proposed method to work. We now proceed to use this framework in a Gaussian scale mixture (GSM) under random scale transformation of the marginals $Y_1, \ldots, Y_q$.

\section{Inferring conditional sign independence in non-Gaussian continuous data using Gaussian scale mixtures}\label{sec:heavy}
 Consider the case where all variables are continuous, but do not necessarily display Gaussian marginal behavior. We formulate the proposed model through a continuous, monotone, \emph{random} transformation function of the marginals $\mathscr F =(f_1, \ldots, f_q)$. Modifying Equation (\ref{eqn:model}), we specify that the transformed data follow a multivariate Gaussian distribution,
\begin{eqnarray}
\mathbf \mathbf\mathscr F(\bY)  | \mathbf \Sigma_{\mathbf G}  &\sim& \mathsf{MN}_{n \times q} (\0, \mathbf I_n, \mathbf \Sigma_{\mathbf G}), \label{eq:mny}
\end{eqnarray}
Two important points to note regarding this formulation are the following:
\begin{enumerate}
\item  In a Bayesian formulation, one can further put priors on each random transformation function, thereby capturing a wide range of marginal behaviors.
\item  \citet{liu09, liu13} showed that for continuous multivariate data, a deterministic monotone transform of the marginals aids interpretability. More specifically, \citet{liu09} showed if the transformation functions $f_1, \ldots f_q$ in Equation (\ref{eq:mny}) are independent and monotone then conditional independence in the transformed data implies conditional independence in the original data. \citet{liu13} relaxed the Gaussianity assumption of Equation (\ref{eq:mny}) to symmetric elliptically contoured distributions. The price one pays for the relaxed assumption is that now it is only possible to infer Kendall's rank correlation \citep{kendall1938new}.
\end{enumerate}
However, not much is known regarding the nature of dependence in the observed data when the transformation functions are random, which is the approach we will take. We start by stating the following definition.
\begin{Def}\label{def:indep}
Two random variables $\zeta_1$ and $\zeta_2$ are said to be conditionally sign independent given $\zeta_3$, if $\mathbb{P}(\zeta_1<0 \cond \zeta_3) = \mathbb P(\zeta_1 <0 \cond \zeta_2, \zeta_3)$; provided these conditional probabilities exist.
\end{Def}
Note that it is only necessary to state the definition in any one direction and the conditional sign independence in the other direction follows readily. We are now ready to state our main result for random scale transformations.
\begin{proposition}\label{prop:markov}
(i) (Conditional sign independence). Consider in Equation (\ref{eq:mny}) the scale transformation $\mathbf \mathbf\mathscr F(\bY) = \Y\D$, where the elements of $\D=\mathrm{diag} (1/d_i)$ are independent with $0<  d_i < \infty$ almost surely with $\int dp(d_i)< \infty$ for $i=1, \ldots, q$. Under the model of Equation (\ref{eq:mny}), $\{\bSigma_\bG^{-1}\}_{\gamma, \nu} =0 \Leftrightarrow \mathbb P(Y_\gamma < 0 | Y_{-\{\gamma, \nu\}}) = \mathbb P(Y_\gamma < 0 | Y_{-\gamma})$. 

\noindent (ii) (Conditional uncorrelatedness). Moreover if $d_i$ are almost surely the same random variable $\tau$ with $\mathbb{E} (\tau^{-1}) < \infty$ then $\{\bSigma_\bG^{-1}\}_{\gamma, \nu} =0 \Leftrightarrow \mathbb{E} (Y_\gamma  | Y_{-\{\gamma, \nu\}}) = \mathbb{E} (Y_\gamma  | Y_{-\gamma}).$
\end{proposition}
The proof is given in \ref{app:markov} Part (i) implies that a missing edge $\{\gamma, \nu\}$ in the graph $\G$ implies the sign of $Y_\gamma$ is independent from that of $Y_\nu$ given the rest of the variables. Admittedly, this result is weaker than conditional independence for Gaussian graphical models (the case where $d_i=1$ for all $i$, a.s.) or, as part (ii) implies, conditional uncorrelatedness for symmetric elliptically contoured distributions (the case where $d_1 = \ldots = d_q = \tau$ a.s. with $\mathbb{E}(\tau^{-1}) < \infty)$. An example of the latter is given by  \citet{finegold11} for the multivariate $t$ distribution. In this case, note that if $\Y\sim t_\nu (\boldsymbol\mu, \Sigma_\bG)$, a multivariate-$t$ distribution with degrees of freedom $\nu$, location vector $\boldsymbol\mu$, and scale matrix $\Sigma_\bG$, then a scale mixture representation is $\Y | \tau, \bSigma_\bG \sim \N(\boldsymbol\mu, \tau\bSigma_\bG), \tau \sim \mathrm{Inv\mbox{-}Gamma}(\nu/2, \nu/2)$. Since the same scale parameter $\tau$ is used for all the margins, conditional uncorrelatedness follows \citep[also proved in Proposition 1 of][]{finegold11}.

This should not come as a surprise, however, since progressively relaxed model assumptions usually come at the cost of progressively weaker statistical conclusions that can be drawn from the model. One cannot expect the relative magnitude among the $Y_i$s to be preserved under different scaling along different marginals. However, the sign of a random variable is independent of its scaling, so long as $0< d_i < \infty$ a.s., providing an intuitive justification of why part (i) of Proposition~\ref{prop:markov} holds.

\subsection{Some examples of continuous marginals in a Gaussian scale mixture}
To further motivate the proposed framework, we now give a few examples of the wide range of marginals we can capture for continuous data in order to infer conditional sign independence.\\
\textbf{Example 1}. \emph{(Power exponential family)}. Consider the (monotone) scale transformation $\mathscr F(\bY) = \bY \D =  \{\by_1/d_1, \ldots, \by_q/d_q\}$ for a $q \times q$ diagonal matrix $\D = \mathrm{diag}(1/d_i)$. Let $p$ be a generic density and consider the Gaussian scale mixture representation
\begin{equation}
p( y_i) = \int_{0}^{\infty} (2 \pi d_i)^{-1/2} \exp(-y_i^2/2d_i) dp(d_i).\label{eq:sm}
\end{equation}
\citet{west1987} showed that the marginal of $ y_i$ is of the form $p(y_i) = k\exp(-| y_i |^b)$ (power-exponential family) if $d_i$ follows a stable distribution with index $b/2$. Since the power-exponential family includes Gaussian ($b=2$) or double-exponential ($b=1$) as special cases, we can make provisions for such marginals.\\
\textbf{Example 2}. \emph{(Generalized hyperbolic family)}. If the mixing distribution in Equation (\ref{eq:sm}) is generalized inverse Gaussian (GIG), the marginals are in the generalized hyperbolic family. This is due to \citet{barn77}  who showed if the mixing distribution is
\begin{eqnarray}
p(d_i) = \frac{(\psi/\chi)^{\lambda/2}}{2K_\lambda(\sqrt{\chi\psi})} d_i^{\lambda-1} \exp\left(-(1/2)(\chi d_i^{-1} + \psi d_i)\right),\label{eq:gig}
\end{eqnarray}
then the marginal is in the generalized hyperbolic family and can be written as
$$
p( y_i) = \frac{(\psi/\chi)^\lambda}{\sqrt{2\pi}K_\lambda(\psi \chi)} \; \!
\times {K_{\lambda - 1/2}\left(\psi \sqrt{\chi + { y_i}^2}\right)} \times {\left(\sqrt{\chi + { y_i^2} / \psi}\right)^{\lambda - 1/2}} \!.
$$
Here $K_\lambda(\cdot)$ is the modified Bessel function of the third kind with index $\lambda$. The domain of the parameters $(\psi, \chi, \lambda)$ and multivariate generalizations are given by 
\citet{barndorff78}. The generalized hyperbolic family includes $t$-distributed marginals as a special case, if each $d_i$ is independent inverse gamma. With the appropriate choice of mixing density on $d_i$, we can have other flexible marginals that are useful, e.g. normal-gamma \citep{grif10} or variance gamma \citep{kotz01}. Table \ref{table:diet} gives some examples of marginal behaviors that we can model, along with corresponding mixing distributions.\\
\textbf{Example 3}.  \emph{(Skewed location-scale family)}. Consider the location-scale transformation $\mathscr F(\bY) =  \{(\by_1 - \mu_1)/d_1, \ldots, (\by_q - \mu_q)/d_q\}$, with the relation $\mu_i = \alpha_i + \beta_i d_i$ for constants $\alpha_i$ and $\beta_i$. In this case, \citet{barn77} showed mixing over $d_i$ with mixing distribution given by Equation (\ref{eq:gig}) gives rise to marginals with asymmetric tails. \emph{This is useful for modeling skewness}. The pure scale transformation is a special case with $\alpha_i = \beta_i = 0$. 

For all the above examples, Metropolis-Hastings samplers can be implemented, enabling practical implementation. While these examples demonstrate the flexibility of the marginal behavior we can model, a fundamental question remains. Given the data, how do we decide what is an appropriate distribution of the scale parameter in a Gaussian scale mixture representation? We prove the following lemma. 
\begin{lemma} \label{prop:bn}
(i) (Polynomially decaying tails). If the tail of the $i$'th marginal $f_i(y_i)$ decays as $ |y_i|^{2\lambda_i -1}$ for some $\lambda_i \leq 0$ as $|y_i| \to\infty$, the mixing distribution of $d_i$ should have tail decaying as $d_i^{\lambda_i-1}$ as $d_i \to \infty$. \\
(ii) (Exponentially decaying tails). If the tail of the $i$'th marginal $f_i(y_i)$ decays as $ |y_i|^{2\lambda_i -1} \exp(-(2\psi_i)^{1/2} |y_i|)$ for some $\lambda_i \in \mathbb{R}, \psi_i>0$, then the mixing distribution of $d_i$ should have tail decaying as $d_i^{\lambda_i-1} \exp(-\psi_i d_i)$ as $d_i \to \infty$.
\end{lemma}
\noindent \emph{Proof.} { (i) This is a consequence of Theorem 6.1 of \citet{barndorff82}. Consider the Gaussian scale mixture $g(x) = \int_{0}^{\infty} \exp(-x^2/2u) (2\pi u)^{-1/2} f(u) du$. \citet{barndorff82} showed if we can write $f(u) \propto u^{\lambda-1} L(u)$ as $u \to \infty$, then  $g(x) \propto |x|^{2\lambda -1} L(x^2)$ as $|x| \to \infty$, where $L(\cdot)$ is a slowly varying function, defined as $\lim_{x\to\infty} L(tx)/L(x)=1$ for any $t\in (0,\infty)$. Since $L(u) \equiv 1$ is slowly varying, we have the desired result.

\noindent (ii) This also follows from the second part of Theorem 6.1 of \citet{barndorff82} by taking $L(u) \equiv 1$.\hfill $\square$}

The above result points to the power of Gaussian scale mixture representation in which the scale can be carefully calibrated to appropriately model the corresponding marginal. In general,  any heavy polynomially decaying tail can be modeled. Tails decaying at exponential rates (e.g., Laplace) can also be modeled. Lemma~\ref{prop:bn} shows that depending on each marginal, one can decide what would be an appropriate mixing density, giving a practical guide to choosing $\D$. For this purpose, plotting marginal q-q plots or histograms will suffice, and one need not be concerned regarding higher order interactions at this point.

Comparing the proposed method to recently proposed techniques, such as the ``alternative multivariate $t$'' \citep{finegold11}, we find two main advantages. First, in our case, the univariate marginals need not all have the same distribution. Our approach includes $t$-distributed marginals of \citet{finegold11} as a special case (if all mixing distributions on the $d_i$s are independent inverse gamma), but is of course, much more flexible. Second, the alternative multivariate-$t$ can only model symmetric tails. However, in our approach, we can make provisions for asymmetric tails using a location-scale mixture, thereby capturing skewness.

\subsection{MCMC procedure for inferring $\bG$} 
We have $ \bY \D =  \{\by_1/d_1, \ldots, \by_q/d_q\} \sim \mathsf{MN}_{n \times q} (\0, \mathbf I_n, \mathbf \Sigma_{\mathbf G})$ for a $q \times q$ diagonal matrix $\D = \mathrm{diag}(1/d_i)$. Let the prior on $\SG$ be $\mathbf \Sigma_{\mathbf G} | \mathbf G, \D \sim \mathrm{HIW_{\mathbf G}} (b, \k\I_q)$. Then, integrating out $\bSigma_\bG$,
\begin{align*}
 \Y\D \cond  \G , \D \sim \HMT{n}{q}{b}{\I_n}{\k\I_q}.
\end{align*}
One can now use suitable mixing distributions on $d_i$ and it is straightforward to perform MCMC to update $\G$ and $\D$, and to obtain samples from the conditional posterior of $(\bSigma_\bG| \bY, \bG, \D)$, as described in Section~\ref{sec:bayesg}. The missing edges in the inferred graph $\G$ points to conditional sign independence among possibly non-Gaussian continuous random variables. It is also possible to integrate out $\D$ completely and formulate the marginal of  $\Y|\G$ up to a constant of proportionality, although we note that the inferred $\D$ provides us knowledge of the marginal behavior through Lemma~\ref{prop:bn}.

\section{Inferring dependence structure across heterogeneous data types}\label{sec:mixed}
In this section we consider the problem of network inference on mixed binary and continuous data. Let our data contain $\bZ\in \{0,1\}^{d}$ discrete and $\bY \in \mathbb{R}^{q}$ continuous variables for the same $n$ samples (with the $d+q$ variables sharing the same dependence structure across all the $n$ samples). A joint model for $\bX=(\bZ,\bY)$ can be specified in terms of the conditionally Gaussian (CG) density of \citet{lauritzen96} as follows:
$$
f(\bx) = f(\bz,\by) = f(\bz) f(\by|\bz) = \exp\left(g_{\bz} + h_{\bz}^T y- \frac{1}{2} \by^T K_{\bz} \by \right).
$$
Define
\begin{eqnarray*}
P_{\bz} = P(\bZ=\bz) = (2\pi)^{q/2} (\det(K_{\bz}))^{-1/2} \exp(g_{\bz} + h_{\bz}^T K_{\bz}^{-1} h_{\bz}/2),\\
\xi_{\bz} = E(\bY | \bZ=\bz) = K_{\bz}^{-1} h_{\bz}, \quad \Sigma_{\bz} = \Var (\bY|\bZ=\bz) = K_{\bz}^{-1},
\end{eqnarray*}
where the conditional distribution of $\bY | \bZ=\bz$ is $\mathrm{N}(\xi_\bz, \Sigma_\bz)$. It is possible to have a fairly general form for the tuple $(g_\bz, h_\bz, K_\bz)$ defining the distribution. Following \citet{cheng13}, we consider a special case of the model
\begin{eqnarray}
\log f(\bz,\by) = \sum_{j=1}^{d} \lambda _j z_j  + \sum_{\substack{{j, k=1}\\ {j>k}}}^{d} \lambda_{jk} z_j z_k + \sum_{\gamma=1}^{q} (\sum_{j=1}^{d} \eta_j^{\gamma} z_j) y_\gamma - \frac{1}{2} \sum_{{ \gamma, \mu=1}}^{q}  y_\gamma k^{\gamma \mu} y_\mu.\label{eq:like}
\end{eqnarray}
Comparing with above, it is clear that we have $g_\bz = \sum_{j=1}^{d} \lambda _j z_j  + \sum_{j>k} \lambda_{jk} z_j z_k; h_\bz^{T} = \sum_{j=1}^{d} \eta_j^{\gamma} z_j$ and $K_\bz=\{k^{\gamma \mu}\}$. Note also that our model is slightly simplified compared to \citet{cheng13}, because $K_\bz$ does  not depend on the discrete variables, the case termed the ``homogeneous model'' by  \citet{lauritzen96}. As pointed out by \citet{cheng13}, this simplified model implies for $j, k \in\{1,\ldots, d\}$ and $\gamma, \mu \in \{1, \ldots, q\}$ that
\begin{eqnarray*}
Z_j \perp Z_k \cond \X \setminus \{Z_j, Z_k\} \Leftrightarrow \lambda_{jk} = 0,\\
Z_j \perp Y_\gamma \cond \bX \setminus \{Z_j, Y_\gamma\} \Leftrightarrow \eta_j^{\gamma} = 0,\\
Y_\mu \perp Y_\gamma \cond \bX \setminus \{Y_\mu , Y_\gamma\} \Leftrightarrow k^{\gamma \mu} = 0.
\end{eqnarray*}
Thus, fitting this model allows one to infer conditional independence relationships across discrete and continuous variables. Note also that the model implies for $j=1, \ldots, d$ and $\gamma = 1, \ldots, q$ the node conditional distributions
\begin{eqnarray}
Z_j  \cond \bX \setminus Z_j &\sim& \mathrm{Binomial} \left(n , \mathrm{logit} \left(\sum_{\substack{k=1 \\ k\neq j }}^{d} \lambda_{jk} Z_k + \sum_{\gamma=1}^{q} \eta_j^{\gamma} Y_\gamma \right) \right), \label{eq:mixed1} \\
Y_\gamma \cond \bX \setminus Y_\gamma &\sim& \mathrm{N} \left(\frac{1}{k^{\gamma \gamma}} \left( \sum_{j=1}^{d} \eta_j^{\gamma} Z_j - \sum_{\substack{\mu =1 \\ \mu \neq \gamma}}^{q} k^{\gamma \mu} Y_\mu\right), \frac{1}{k^{\gamma \gamma}} \right) \label{eq:mixed2},
\end{eqnarray}
where $\mathrm{logit}(\psi) = (1+ \exp(-\psi))^{-1}$ for $\psi \in \mathbb{R}$. In the case of purely discrete or purely continuous data, the above conditional relationships correspond to a joint Ising distribution for discrete data and a joint multivariate Gaussian distribution for continuous data, respectively \citep{lauritzen96}. Directly maximizing the joint log likelihood in Equation (\ref{eq:like}) is known to be difficult \citep{lee2015learning, cheng13}. Thus, following the neighborhood selection approach of \citet{meinhausen06}, existing works for pure discrete data fit penalized logistic regressions for the discrete part \citep[e.g., ][]{ravikumar10} and penalized Gaussian regressions for the  continuous part \citep[e.g., ][]{friedman08} in high-dimensional settings to maximize the node conditional likelihoods (or pseudolikelihoods) of Equations (\ref{eq:mixed1}-\ref{eq:mixed2}). Building on these, \citet{cheng13} devised an alternating algorithm to simultaneously fit both types of regressions for mixed data. However, a rather surprising fact is that the logistic distribution can be written as a Gaussian location-scale mixture as well. We now show this allows a direct characterization of the joint density of $(\Z,\Y)$ as a multivariate normal, conditional on mixing P\'{o}lya-Gamma variables for the discrete parts. To begin, note that if $U \sim \mathrm{Binomial} (n, \mathrm{logit}(\psi))$ then \citet{polson13pg} demonstrated the following location-scale mixture representation: 
\begin{eqnarray*}
\left(U - \frac{n}{2}\right) \cond \omega &\sim&  \mathrm{N}(\omega\psi, \omega); \quad
\omega \sim \mathrm{PG}(n,0),
\end{eqnarray*}
where $\mathrm{PG} (n,0)$ denotes a P\'{o}lya-Gamma random variable, which can be expressed as an infinite weighted sum of Gamma random variables. Its density and moments are given by \citet{polson13pg} and an efficient sampler is available in the R package \texttt{BayesLogit} \citep{polson2012package}. Introducing latent P\'{o}lya-Gamma variables, Equations (\ref{eq:mixed1}) and (\ref{eq:mixed2}) become
\begin{eqnarray}
\left(Z_j - \frac{n}{2}\right) \cond \omega_j, \bX \setminus Z_j &\sim& \mathrm{N} \left(\omega_j\left(\sum_{\substack {k=1 \\ k\neq j }}^{d} \lambda_{jk} Z_k + \sum_{\gamma=1}^{q} \eta_j^{\gamma} Y_\gamma\right) , {\omega_j}\right), \label{eq:mm1} \\
\omega_j &\stackrel{i.i.d}\sim& \mathrm{PG}(n,0),\nonumber\\
Y_\gamma \cond \bX \setminus Y_\gamma &\sim& \mathrm{N} \left(\frac{1}{k^{\gamma \gamma}} \left( \sum_{j=1}^{d} \eta_j^{\gamma} Z_j - \sum_{\substack{\mu =1 \\ \mu \neq \gamma}}^{q} k^{\gamma \mu} Y_\mu\right), \frac{1}{k^{\gamma \gamma}} \right). \label{eq:mm2}
\end{eqnarray}
One can now see from Equations (\ref{eq:mm1}) and (\ref{eq:mm2}) that all the $(d+q)$ node conditional distributions of one variable given the rest follow univariate normal distributions. By properties of multivariate normal, the joint distribution of the variables  $(\Z,\Y)$ given $\boldsymbol \omega = (\omega_1, \ldots, \omega_d)$ must also correspond to a multivariate normal that will preserve these conditional means and variances \citep[see, e.g.,][]{khatri76}. Thus, define the transformed data
\begin{eqnarray}
\tilde \X = \left(Z_1-n/2, \ldots, Z_d-n/2, Y_1, \ldots, Y_q\right) \cond \boldsymbol \omega &\sim& \mathsf{MN}_{n \times (d+q)} (\0, \bI_n, \bSigma), \label{eq:mixed3}\\
\omega_j &\stackrel{i.i.d}\sim& \mathrm{PG}(n,0), \quad \text{for} \quad j=1, \ldots, d. \label{eq:mixed4}
\end{eqnarray}
Define $\lambda_{ii} = 1/\omega_i$. Then, the $(d+q) \times (d+q)$ symmetric $\bSigma^{-1}$ is given by
\begin{equation*}
\bSigma^{-1} = \left(
\begin{array} {cccccc}
\lambda_{11}  &\ldots &-\lambda_{1d} &-\eta_{1}^{1}& \ldots &-\eta_{1}^{q} \\
\vdots &\ddots &\vdots &\vdots &\ddots &\vdots\\
-\lambda_{d1}  & \ldots& \lambda_{dd} & -\eta_{d}^{1}& \ldots& -\eta_{d}^{q} \\
-\eta_1^{1}  &\ldots& -\eta_d^{1} & k^{11}  & \ldots &k^{1q}\\
\vdots &\ddots &\vdots &\vdots &\ddots &\vdots\\
-\eta_1^{q}  &\ldots & -\eta_d^{q} & k^{q1}  & \ldots &k^{qq}\\
\end{array} \right). 
\end{equation*}
The $\omega_i$ terms are independent and one can easily verify that $\int dp(\omega_i)  < \infty$ when $\omega_i \sim \mathrm{PG} (n,0)$. Note that an inverse Wishart prior on $\bSigma$ is not sensible any more because that will not induce inverse P\'{o}lya-Gamma priors on $(\lambda_{11}, \ldots, \lambda_{dd})$. Thus in order to model this inverse covariance matrix, we follow the idea introduced by \citet{wong2003}, who decouple the modeling for the diagonal and off-diagonal elements. Write 
$$
\bOmega= \bSigma^{-1} = \bTheta \bGamma \bTheta,
$$
where $\bTheta$ is a $(d+q)$ diagonal matrix with $i$th diagonal entry $\Theta_i =  \sqrt{\bOmega_{ii}}$ and and $\bGamma$ is related to $\bOmega$ as $\bGamma_{ij} = - \bOmega_{ij}/\sqrt{\bOmega_{ii}\bOmega_{jj}}$, i.e., the entries of $\bGamma$ are the negative of the partial correlation matrix, with ones on the diagonal \citep{wong2003}. Then, we parameterize 
\begin{eqnarray}
(\Theta_{1}^2, \ldots, \Theta_{d}^2) = ({\lambda_{11}} , \ldots, {\lambda_{dd}}) &\sim& 1/\mathrm{PG} (n,0), \label{eqn:mcmc1}\\
(\Theta^2_{d+1}, \ldots, \Theta^2_{d+q}) = (k^{11}, \ldots, k^{qq}) &\sim& 1/\mathrm{Inv\mbox{-}Gamma} (\alpha, \beta), \label{eqn:mcmc2}
\end{eqnarray}
where all random variables are distributed independently and $\alpha, \beta$ are hyperparameters. We follow the same prior specification on the entries on $\bGamma$ as \citet{wong2003}, which enables a sparse estimation of $\bGamma$. Thus, our parameterization differs from that of \citet{wong2003} only for the entries $(\Theta_{1}^2, \ldots, \Theta_{d}^2)$ where they use Gamma priors, and we need to use inverted P\'{o}lya-Gamma priors. We conjecture that using the representation of P\'{o}lya-Gamma random variable as an infinite weighted sum of gamma random variables, it might be possible to characterize the induced distribution on $\bSigma^{-1}$ more explicitly, although we have not pursued this. In any case, with this modification, one can employ the same MCMC sampling procedure as in  \citet{wong2003} in order to iteratively update $(\bTheta_i | \tilde \X , \bTheta_{-i}, \bGamma)$ and  $(\bGamma_{ij} |  \tilde \X , \bTheta, \bGamma_{-\{ij\}})$. Conditional independence holds according to off-diagonal zeros in inferred $\bGamma$, between the discrete-discrete, continuous-continuous or discrete-continuous random variables. Further note that we have assumed the continuous part of the data follows multivariate Gaussian distribution. An application of Proposition~\ref{prop:markov} shows that non-normal marginals can be modeled by appropriate choices of scale distributions for each marginal $Y_1, \ldots, Y_q$ and one would still be able to infer conditional sign independence. Contrast this with the framework of \citet{cheng13}, which is not equipped to handle non-normal marginals.

Following the well-known latent variable technique of \citet{albert93} for probit models, the existing literature for Bayesian modeling of mixed data introduces a latent continuous counterpart for the observed discrete data for which posterior sampling is feasible \citep{pitt06, dobra11}. Conditional independence is then inferred among the observed and latent continuous variables. Unfortunately, there is no direct characterization of the conditional independence relationship between the observed discrete data and their latent counterpart \citep{pitt06}. Our approach overcomes this difficulty through a direct scale transformation and we can infer dependence relationship directly at the level of the observed data.

\section{Simulation study} \label{sec:sim}
We performed simulation experiments comparing the proposed method with competing approaches. We present the results for continuous non-Gaussian data and mixed discrete-continuous data in Sections \ref{sec:simcont} and \ref{sec:simmixed} respectively.

\subsection{Non-normal continuous data}\label{sec:simcont}

\begin{figure}[!t]
\begin{center}
\includegraphics[height=10cm,width=18cm]{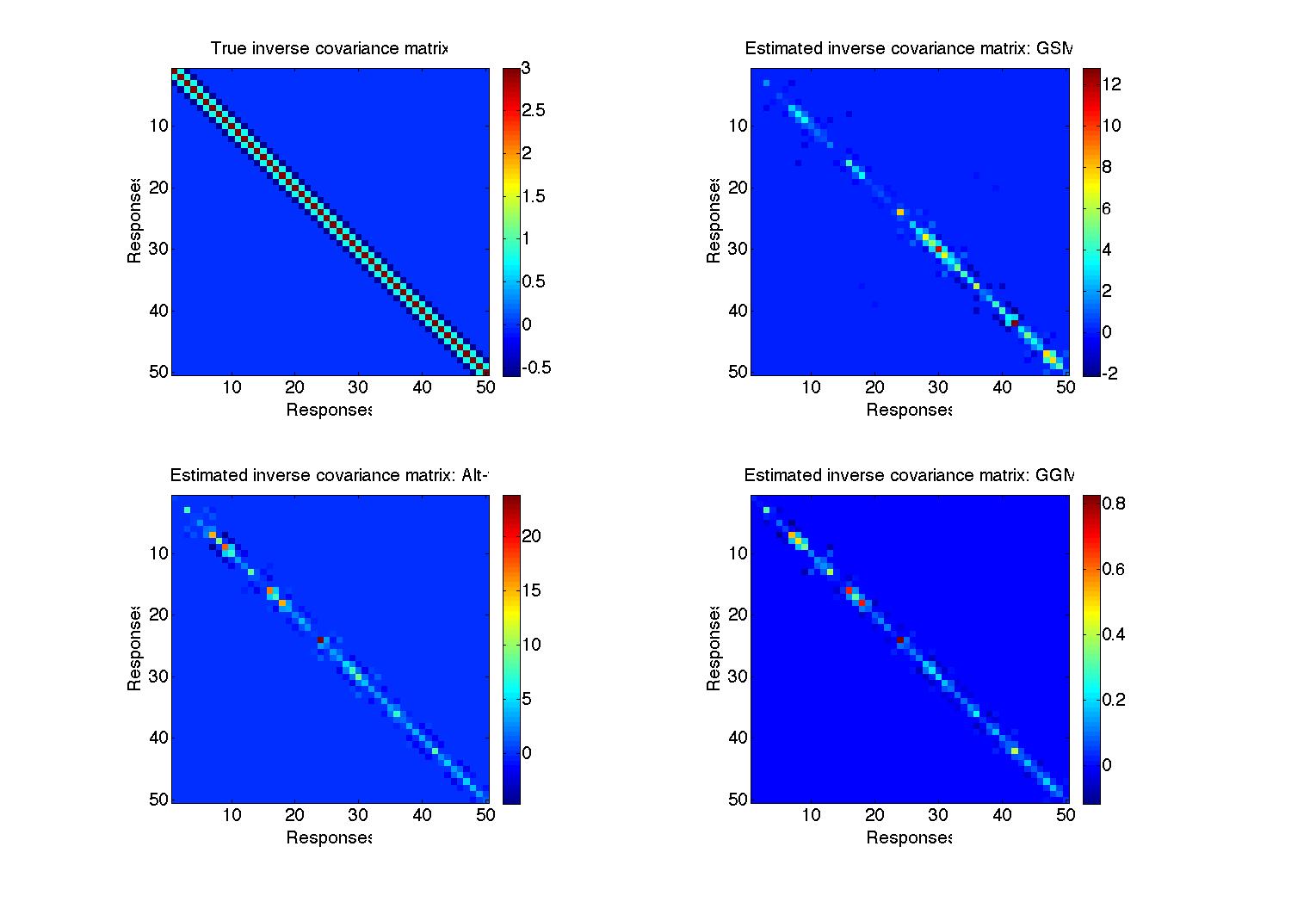}
\end{center}
\caption{True and estimated $\Sigma^{-1}$ for continuous non-normal data. Clockwise from top left: true, estimated by proposed method using Gaussian scale mixtures (GSM), by Gaussian graphical model (GGM) and by alternative multivariate-$t$ (Alt-t).\label{fig:inv}}
\end{figure}

We chose $n=100$ and $q=50$. We then simulated data according to the true inverse covariance matrix shown on the top left of Figure~\ref{fig:inv}. The true $\bSigma^{-1}$ is a symmetric banded diagonal matrix with diagonal elements equal to $v=3$, the first sub-diagonal $=0.25v = 0.75$ and the second subdiagonal $=-0.2v = -0.6$, the rest of the elements being zero. Thus, the true inverse covariance matrix is sparse and there are both positive and negative partial correlations present. Positive definiteness for the resulting matrix can be easily verified using the diagonal dominance property. We simulate data as $\bY \sim \mathrm{MN} (0, \bI_n, \bSigma)\cdot \D$. Where $\D=\mathrm{diag}(1/d_i)$ is a diagonal matrix with $d_i \sim \mathrm{Exponential} (\mathrm{mean}=10)$ for $i=1, \ldots, 25$ and $d_i \sim \mathrm{Inv\mbox{-}Gamma} (\mathrm{shape=}3, \mathrm{scale=} 10)$ for $i=26, \ldots, 50$. Thus, the first 25 marginals in the observed data have double-exponential distribution while the remaining 25 have polynomially decaying $t$-distribution (refer to Table~\ref{table:diet}).

For this data, we compared four approaches: the proposed method based on Gaussian scale mixtures (GSM), alternative multivariate-$t$ (Alt-t) of \citet{finegold11}, a sparse Bayesian Gaussian graphical model (GGM) as described in Section~\ref{sec:bayesg} and the Gaussian copula graphical model (GCGM) of \citet{pitt06}. We implemented the first three methods in MATLAB and for GCGM we used the implementation in the R package \texttt{BDgraph} by \citet{mohammadi15}. GGM is implemented according to Equations (\ref{eq:gamma}-\ref{eqn:model}). For hyperparameters we used $b=10, \rho=0.5$ and prior weight $w_{uv} = 0.1$ for all edges in this example, but performed sensitivity analysis to ensure the choice of hyperparameters do not have a large effect on results. To implement Alt-$t$, we further put independent $\mathrm{Inv\mbox{-}Gamma} (2, 7)$ prior on all $d_i$. To implement GSM, we put independent $\mathrm{Exponential}(5)$ on the first 25 and $\mathrm{Inv\mbox{-}Gamma} (2, 7)$  on the rest. Results appear to be stable over a range of hyperparameter values. We used 50,000 MCMC iterations with a burn-in period of 20,000 iterations for all methods. Figure~\ref{fig:inv} shows the true and estimated $\bSigma^{-1}$ for the first three methods (see Figure~\ref{fig:supp_heavy} in the supplement for the estimate of GCGM). An interesting observation is the scale next to each panel. It appears the Gaussian graphical model deals with different scaling across different marginals, for which it is a misspecified model, by heavily shrinking all entries of the resultant estimate of $\bSigma^{-1}$. On the other hand, the alternative-$t$, which expects polynomially decaying $t$ marginals along all coordinates, appears to inflate the absolute values of some of the resulting estimates compared to the proposed method. Nevertheless, we remind the reader that the values of estimated $\bSigma^{-1}$ are not directly comparable across the three methods, although their signs are. Table~\ref{tab:inv} reports the detection of correct sign of the elements of true $\bSigma^{-1}$ (zero, positive or negative) by the three competing methods. The ratio of estimated vs. true is shown the table, with the actual counts in parentheses. A ratio close to 1 indicates superior performance by a method. It is clear the proposed approach has the best performance in all three categories (detection of true zero as zero, and similarly for positive and negative elements). Alt-t has the second best performance and GCGM actually performs the worst in this setting, by underestimating the number of true zeros and overestimating both the numbers of positive and negative elements. For this data, we also tried non-Bayesian graphical lasso method, but it failed to converge after 5,000 iterations and we do not have numeric values to report. We also experimented with other sparse structures of the true $\bSigma^{-1}$. We considered structured cases, such as top left $5\times 5$ off-diagonal block non-zero (half of them positive, the other half negative), rest off-diagonals zero; and unstructured cases, such as randomly selected 5\% elements positive, 5\% negative, rest 0, subject to the condition that this corresponds to a valid decomposable graph. Positive definiteness was ensured by diagonal dominance. The finding that the proposed method displays superior performance in sign detection remains robust. 

\subsection{Mixed binary and continuous data} \label{sec:simmixed}

\begin{figure}[!t]
\begin{center}
\includegraphics[height=7cm,width=18cm]{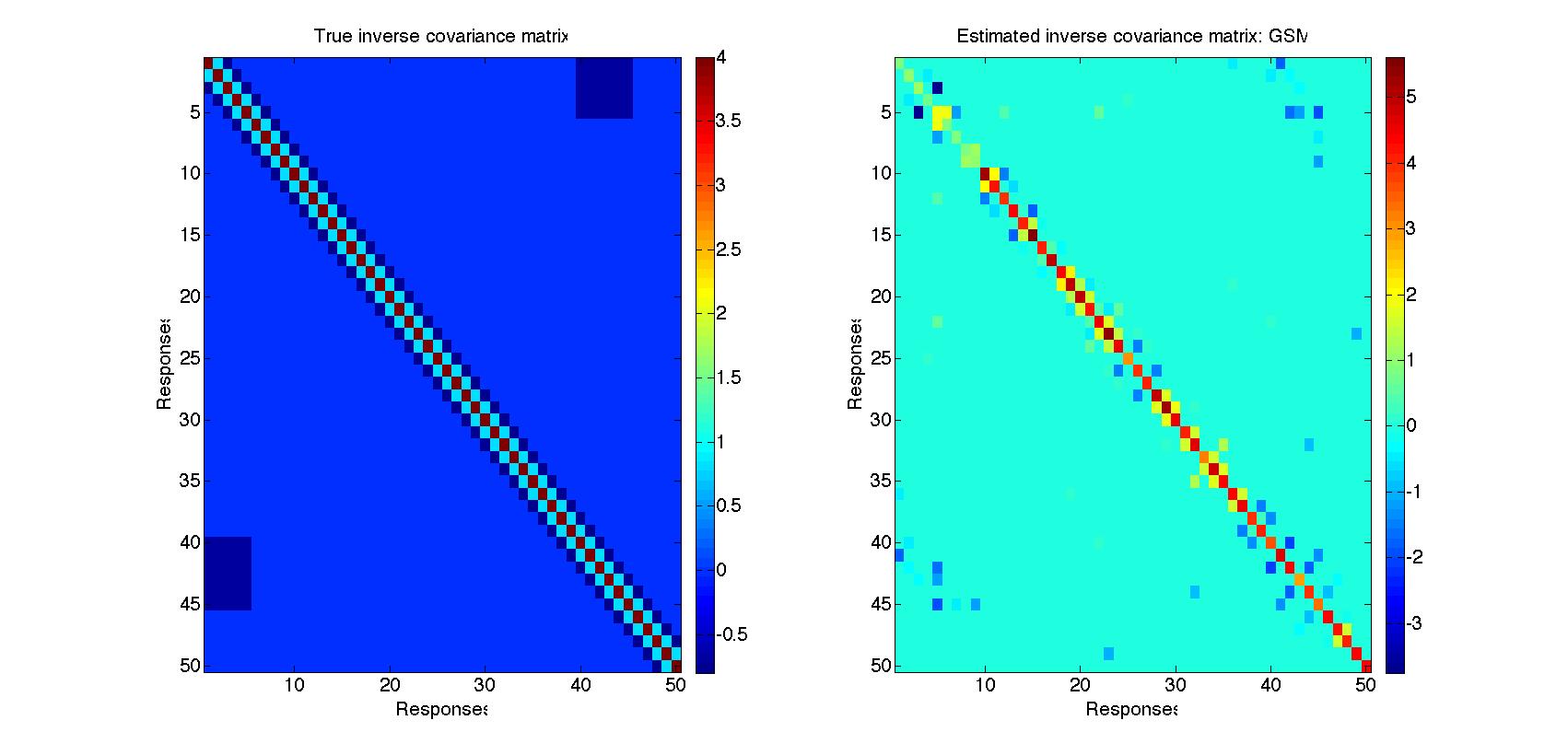}
\end{center}
\caption{True and estimated $\Sigma^{-1}$ for mixed discrete and continuous data. Left: true, right: estimated by the proposed Gaussian scale mixture (GSM) method. \label{fig:mixed}}
\end{figure}

Here we chose $n=100, d=9$ and $q=41$. That is, we considered a total of 50 variables, the first 9 of them discrete and the remaining 41 continuous and there are 100 observations for each variable. The true inverse covariance matrix is shown in the top panel of Figure~\ref{fig:mixed}. The true $\bSigma^{-1}$ is a symmetric banded diagonal matrix with diagonal elements equal to $v=4$, the first sub-diagonal $=0.2v = 0.8$ and the second subdiagonal $=-0.2v = -0.8$. In addition, we wanted to see if the method can successfully capture dependence between discrete and continuous random variables. Thus, we set $\bSigma^{-1}_{1:5, 40:45} = \bSigma^{-1}_{40:45, 1:5} = -0.7$, introducing negative dependence. The mixed discrete and continuous data were then simulated according to the Equations (\ref{eq:mixed3}-\ref{eq:mixed4}). In order to create discrete observations, we rounded each entry of the first 9 columns to the nearest integer. 

For estimation purposes, we comparde the performance of GSM and GCGM. As in the previous subsection, we used native MATLAB implementation of GSM and the implementation in the package \texttt{BDgraph} for GCGM. To implement GSM, we used the parameterization in Equations (\ref{eqn:mcmc1}-\ref{eqn:mcmc2}). We simulated the required $\mathrm{PG}(n,0)$ random variables using the \texttt{Bayeslogit} package. For the hyperparameters, we used $\alpha=\beta=1/2$ which appeared to work well in practice. As before we used 50,000 MCMC iterations and a burn-in period of 20,000 iterations and monitored the log-likelihood to ensure convergence. The estimated $\bSigma^{-1}$ by GSM is shown in the right panel of Figure~\ref{fig:mixed} (see Figure \ref{fig:supp_mixed} in the supplement for the estimate of GCGM). 
The performance of GSM and GCGM in terms of capturing conditional sign dependence is reported in Table~\ref{tab:mixed}. Note that the alternative multivariate-$t$ and Gaussian graphical models are not suited for comparisons over mixed discrete-continuous data. Although GCGM of \citet{pitt06} can work with mixed discrete and continuous data, the interpretation of their estimated covariance matrix, which uses a latent continuous counterpart for the discrete variables, differs from ours which uses no such latent variable representation, other than the mixing P\'{o}lya-Gamma scale parameter. 
Nevertheless, it appears from Table~\ref{tab:mixed} that GCGM does a poor job compared to GSM. It underestimates the number of zeros and overestimates the number of both positive and negative entries. In other words, the estimate is not as sparse as it should be, which is also apparent from Figure~\ref{fig:supp_mixed}. This finding of the behavior of GCGM is also consistent with Section~\ref{sec:simcont}, where it tends to produce a less sparse estimate compared to the other methods. Recall that both our approach (GSM) and GCGM can work with non-Gaussian distributions for the continuous data. Thus, although the data in this simulation uses normal marginals for continuous components, we experimented with non-normal marginals and the results remain quite robust. 

\section{Analysis of glioblastoma multiforme data} \label{sec:real}
Our data consists of continuous expression levels and mutation status for 49 genes that overlap with the three critical signaling pathways - the RTK/PI3K signaling pathway, the p53 signaling pathway, and the Rb signaling pathway, which are known to be involved in migration, survival and apoptosis progression of cell cycles in GBM \citep{furnari2007malignant}. Of these 49 genes, 20 did not not show evidence of mutation in any location. Thus, our data consists of $q=49$ gene expressions and $d=29$ binary mutations for $n=103$ glioblastoma multiforme (GBM) patients. The raw data are publicly available through the Cancer Genome Atlas (TCGA) data portal (\url{http://tcga-data.nci.nih.gov/tcga/}). 
We standardize the continuous components by subtracting the mean and dividing by the standard deviation. In Figure~\ref{fig:tcga}, we provided an illustration of non-normal marginals in the continuous components by plotting the expression levels for AKT3 and CDK4 genes. These non-normal features are preserved under standardization. The complete list of genes whose expression levels and mutation status we consider is given in Supplementary Table~\ref{tab:supp_genes}.

\begin{figure}[!t]
\begin{center}
\includegraphics[height=18cm,width=18cm]{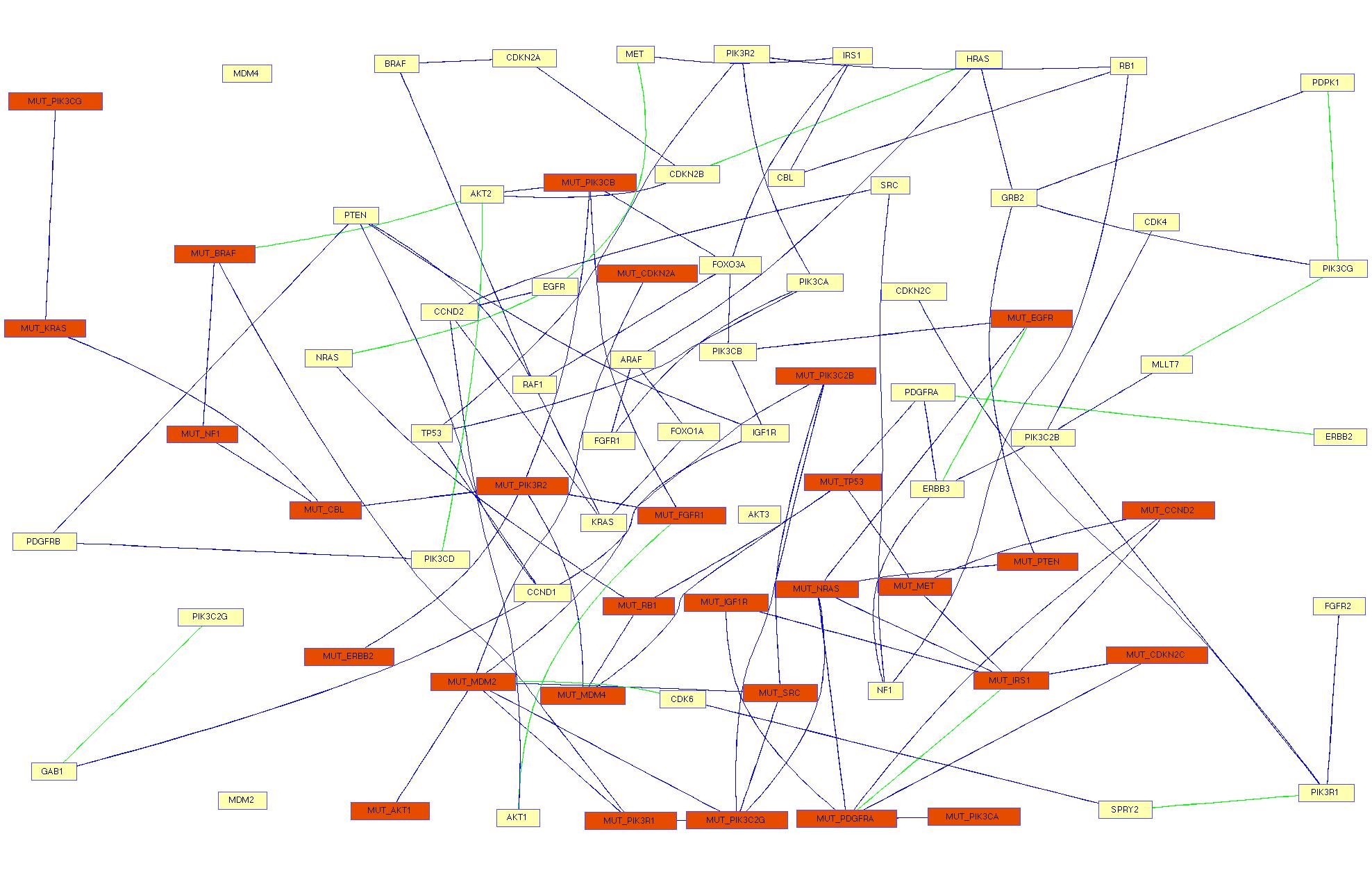}
\end{center}
\caption{The estimated conditional sign dependence network on glioblastoma multiforme mutation and expression data. A node with a subscript ``MUT\_'' denotes a binary mutation (in red). Otherwise it denotes a continuous valued gene expression (in yellow). A blue edge corresponds to a negative estimated inverse covariance entry, green corresponds to positive.\label{fig:real}}
\end{figure}

We illustrate in Figure~\ref{fig:real} the conditional sign dependence network obtained by the proposed Gaussian scale mixture (GSM) method. Each connection represents a non-zero entry in the estimated inverse covariance matrix. Nodes with high connectivity appear closer to the center of the figure and those with lower degrees of connectivity are closer to the edges. A red colored node with a subscript ``MUT\_'' denotes in the figure that the node corresponds to a binary mutation in a given gene; and a yellow colored node represents a continuous valued expression level. Several mutations show a high degree of negative association to other mutations and to expression levels of other genes. This includes the mutations in TP53 (negatively associated with mutations in MDM4, RB1, MET and to the expression level of PDGFRA), mutations in FGFR1 (negatively associated with mutations in PIK3R2, PIK3CB and positively to the expression levels of AKT1), mutations in PIK3R2 (negatively associated to mutations in FGFR1, ERBB2 and PIK3CB). Expression levels of IGF1R shows a high degree of connectivity (negatively to expression levels of PIK3CB, PTEN, CCND1). 

On the other hand, some other expression levels appear isolated and do not appear to be connected to the other mutations and expressions under consideration. These include the expression levels of the MDM family (MDM2 and MDM4). It is interesting to note however that the mutations in the MDM family of genes are connected to other nodes, suggesting that this mutation acts by changing the expression levels of other genes (i.e., exhibits a \emph{trans} effect). The influence of mutations in TP53 for GBM has been known to affect the prognosis \citep{shiraishi2002influence} and its reactivation via an MDM inhibitor has been observed \citep{costa2013human}, suggesting an interaction. Our analysis is in accordance with known pathway interactions in GBM \citep[e.g., compare with Figure 4A of ][]{brennan2013} and uncovers several new associations via joint analysis of binary and continuous valued data.

\section{Conclusions} \label{sec:conc}
We proposed an approach based on Gaussian scale mixtures that is capable of handling the problem network inference in presence of non-normal marginals and mixed discrete and continuous random variables in a unified framework. We introduced the concept of conditional sign independence and showed that it is possible to infer this based on the proposed method.  By this measure, we showed by simulations that the proposed method performs better than alternatives such as copula Gaussian graphical models. 

Some natural extensions of the proposed framework can be considered as future work. Prominent among them is the extension of the mixed binary/continuous framework in Section~\ref{sec:mixed} to the mixed binary/ordinal/continuous case. In this case, the discrete variables would follow a multi-category logistic model instead of just two, and one may proceed using the framework of \citet{polson13pg} for multiple categories. Although for the purpose of this paper we are interested in Bayesian techniques, a scale mixture approach lends itself naturally to expectation-maximization (E-M) algorithms for maximizing likelihoods. If one is interested in estimating the inverse covariance matrix in a penalized likelihood framework, one can use our proposed framework where in the E-step instead of sampling $\bTheta$, one would substitute its conditional expectation given the rest, and simulation of $\bGamma$ would be replaced by a penalized Gaussian likelihood maximization step, which is usually quite simple. For the special case of alternative multivariate-$t$, the E-M scheme was discussed by \citet{finegold11}. The current framework shows it is applicable more broadly, as long as one is able to compute the posterior expectations. This is especially promising for the case of mixed binary and continuous data, since \citet{polson13pg} provide very simple formulas for the expectation of P\'{o}lya-Gamma random variables. Thus, even in the non-Bayesian case, our proposed framework points to a possible alternative latent variable framework for implementing E-M to find the mle and it would be interesting to compare its performance to the pseudolikelihood approaches of  \citet{cheng13} or \citet{lee2015learning}.
\section*{Sumpplementary material}
The supplementary file contains 
additional figures and tables referenced in Sections~\ref{sec:sim} and~\ref{sec:real}. 
\newcommand{\Appendix}{\appendix\def\thesection{Appendix~\Alph{section}}\def\thesubsection{\Alph{section}.\arabic{subsection}}}
\begin{appendix}
\Appendix
\renewcommand{\theequation}{A.\arabic{equation}}
\renewcommand{\thesubsection}{A.\arabic{subsection}}
\renewcommand{\thesection}{Appendix \Alph{section}.}
\setcounter{equation}{0}
\baselineskip=\gnat
\section{Proof of Proposition \ref{prop:markov}}\label{app:markov}
We have
\begin{eqnarray*}
\mathbf \mathbf\mathscr F(\bY)  | \mathbf \Sigma_{\mathbf G}  &\sim& \mathsf{MN}_{n \times q} (\0, \mathbf I_n, \mathbf \Sigma_{\mathbf G}), 
\end{eqnarray*}
where  we define the $n \times q$ matrix $\widetilde \bY = \mathbf \mathbf\mathscr F(\bY) = \{f_1(\by_1), \ldots, f_q(\by_q)\}$, with each $\by_i$ being a column vector of length $n$. Consider the scale transformation $\bY \D =  \{\by_1/d_1, \ldots, \by_q/d_q\}$ for a $q \times q$ diagonal matrix $\D = \mathrm{diag}(1/d_i)$. This gives
\begin{eqnarray*}
\mathbf \bY\D  |  \mathbf \Sigma_{\mathbf G}, \D  &\sim& \mathsf{MN}_{n \times q} (\0, \mathbf I_n, \mathbf \Sigma_{\mathbf G}) \label{eq:amny}.
\end{eqnarray*}
Let $\SG^{-1} =K$ and let 
$$
K_{\{\gamma, \nu\}} =  \left(\begin{array} {cc} k_{\gamma\gamma} & k_{\gamma \nu}\\ k_{\gamma \nu} & k_{\nu \nu} \end{array} \right).
$$
Then,
\begin{eqnarray*}
p(Y_\gamma, Y_\nu | \Y_{-\{\gamma, \nu\}} , \D) &=& (2\pi)^{-1}\mathrm{det} (K_{\{\gamma, \nu\}})^{1/2}\\
&&  \times \exp\left(-\frac{1}{2} \left(\frac{Y_\gamma}{d_\gamma} - \mu_{\gamma,\D}; \frac{Y_\nu}{d_\nu} - \mu_{\nu,\D}\right)^T K_{\{\gamma, \nu\}} \left(\frac{Y_\gamma}{d_\gamma} - \mu_{\gamma,\D}; \frac{Y_\nu}{d_\nu} - \mu_{\nu,\D}\right)\right),
\end{eqnarray*}
where $\mu_{\gamma,\D}$ is the mean of $(Y_\gamma /d_\gamma)$ given $\D$ and $Y_{-\{\gamma, \nu\}}$ and similarly for $\mu_{\nu,\D}$.  First assume $k_{\gamma\nu}=0$. Then we have
\begin{eqnarray*}
p(Y_\gamma, Y_\nu | \Y_{-\{\gamma, \nu\}} , \D) &=& (2\pi)^{-1/2} k_{\gamma\gamma}^{1/2}\exp\left(-\frac{1}{2} \left(\frac{Y_\gamma}{d_\gamma} - \mu_{\gamma,\D}\right)^T k_{\gamma \gamma} \left(\frac{Y_\gamma}{d_\gamma} - \mu_{\gamma,\D}\right) \right) \\
&\times& (2\pi)^{-1/2} k_{\nu\nu}^{1/2}\exp\left(-\frac{1}{2} \left(\frac{Y_\nu}{d_\nu} - \mu_{\nu,\D}\right)^T k_{\nu\nu} \left(\frac{Y_\nu}{d_\nu} - \mu_{\nu,\D}\right) \right) .
\end{eqnarray*}
where we can deduce from Proposition C.5 of \citet{lauritzen96} that
$$
\mu_{\gamma,\D} = - \frac{1 }{ k_{\gamma \gamma}}\sum _{\xi \neq \gamma, \nu} \frac{k_{\gamma\xi} Y_\xi}{d_\xi}; \mu_{\nu,\D} = - \frac{1 }{ k_{\nu \nu}}\sum _{\xi \neq \gamma, \nu} \frac{k_{\nu\xi} Y_\xi}{d_\xi}.
$$
We also have
\begin{eqnarray*}
p(Y_\gamma | \Y_{-\gamma}, \D ) &=& (2\pi)^{-1/2} k_{\gamma\gamma}^{1/2}\exp\left(-\frac{1}{2} \left(\frac{Y_\gamma}{d_\gamma} - \tilde \mu_{\gamma,\D}\right)^T k_{\gamma \gamma}\left(\frac{Y_\gamma}{d_\gamma} - \tilde\mu_{\gamma,\D}\right) \right),
\end{eqnarray*}
where $
\tilde \mu_{\gamma,\D} = - (1 / k_{\gamma \gamma}) \sum _{\xi \neq \gamma} (k_{\gamma\xi} Y_\xi)/d_\xi.$
So under the restriction $k_{\gamma \nu}=0$ we have $\mu_{\gamma,\D} = \tilde \mu_{\gamma,\D}$ and also $\mu_{\nu,\D}= \tilde \mu_{\nu,\D}$. Thus,
$$
p(Y_\gamma, Y_\mu | \Y_{-\{\gamma, \mu\}} , \D) = p(Y_\gamma | \Y_{-\gamma}, \D ) p(Y_\nu | \Y_{-\nu} , \D).
$$
Clearly, conditional independence does not hold after integrating out $\D$. Conditional uncorrelatedness also does not hold unless all $d_i$ with $i=1, \ldots , q$ are the same random variable. To see this note the following:
\begin{eqnarray*}
\mathbb{E} [Y_\gamma \cond \Y_{-\gamma}] &=& \mathbb{E} _{\D \cond \Y_{-\gamma}} \left[ \mathbb{E} [ Y_\gamma \cond \Y_{-\gamma}, \D] \right ],\\
\mathbb{E} [Y_\gamma \cond \Y_{-\{\gamma,\nu\}}] &=& \mathbb{E} _{\D \cond \Y_{-\{\gamma,\nu\}}} \left[ \mathbb{E} [ Y_\gamma \cond \Y_{-\{\gamma,\nu\}}, \D] \right ].
\end{eqnarray*}
The two inner conditional expectations on the right hand sides are equal, the value being
$$
\mathbb{E} [ Y_\gamma \cond \Y_{-\gamma}, \D] = \mathbb{E} [ Y_\gamma \cond \Y_{-\{\gamma,\nu\}}, \D] =- \frac{d_\gamma }{ k_{\gamma \gamma}}\sum _{\xi \neq \gamma, \nu} \frac{k_{\gamma\xi} Y_\xi}{d_\xi},
$$
but the conditional densities of $(\D \cond Y_{-\gamma})$ and $(\D \cond Y_{-\{\gamma,\nu\}})$ are not equal. Hence the two resultant left hand sides are not equal after computing the outer expectations. A special case is of course when $\D$ is just a single random variable used for all margins. Then it is easy to see the inner expectations on the right hand sides are constant with respect to $\D$ and conditional uncorrelatedness follows, completing the proof of part (ii) \citep[see also Proposition 1 of][]{finegold11}. But note that we still have
\begin{eqnarray*}
\mathbb P(Y_\gamma < 0 | Y_{-\{\gamma, \nu\}} )&=& \mathbb{E}_{\D \cond Y_{-\{\gamma, \nu\}}} \left[\mathbb P\left(\frac{Y_\gamma}{d_\gamma }< 0 | Y_{-\{\gamma, \nu\}}, \D \right)\right]\\
&=& \mathbb{E}_{\D \cond Y_{-\{\gamma, \nu\}}} \left[ \Phi \left(\frac{1 }{ k_{\gamma \gamma}}\sum _{\xi \neq \gamma, \nu} \frac{k_{\gamma\xi} Y_\xi}{d_\xi}\right) \right]\\
&=& \mathbb{E}_{\D_{-\{\gamma, \nu\}} \cond Y_{-\{\gamma, \nu\}}} \left[ \Phi \left(\frac{1 }{ k_{\gamma \gamma}}\sum _{\xi \neq \gamma, \nu} \frac{k_{\gamma\xi} Y_\xi}{d_\xi}\right) \right]\\
&=& \mathbb{E}_{\D_{-\{\gamma, \nu\} \cup \nu} \cond Y_{-\{\gamma, \nu\}\cup \nu}} \left[ \Phi \left(\frac{1 }{ k_{\gamma \gamma}}\sum _{\xi \neq \gamma, \nu} \frac{k_{\gamma\xi} Y_\xi}{d_\xi}\right) \right]\\
&=& \mathbb P(Y_\gamma < 0 | Y_{-\gamma} ).
\end{eqnarray*}
The third display is true since the integrand does not depend on $d_\gamma$ and $d_\nu$ and the posteriors of $d_i | Y_i$ are independent for $i=1, \ldots , q$, 
giving the desired result in part (i) when $0< d_i <\infty$ with $\int dp(d_i) < \infty$ ensuring the existence of the integrals.
\end{appendix}

\baselineskip=13pt
\bibliographystyle{biom} \bibliography{bayes,highdim,reff,tree,treeTest}

\baselineskip=20pt
\clearpage\pagebreak\newpage

\clearpage\pagebreak\newpage

\clearpage\pagebreak\newpage

\clearpage\pagebreak\newpage

\clearpage\pagebreak\newpage
\begin{table}[t]
\centering
\footnotesize
\begin{tabular}{cc|cc}
\hline
\hline
Marginal for $ y_i$ & Mixing distribution of $d_i$ & Marginal for $ y_i$ & Mixing distribution of $d_i$  \\
(Power-exponential family)   & (Stable family) & (Generalized-hyperbolic family) & (GIG family)\\
\hline
Double-exponential & Exponential & Cauchy (or Student-$t$) & Inverse gamma\\
 Gaussian & Degenerate (constant) & Logistic & P\'{o}lya-gamma\\
\hline
\end{tabular}
\caption{Some illustrations of non-Gaussian marginals in a Gaussian scale mixture with corresponding mixing distribution of the scale parameter.\label{table:diet}}
\end{table}

\begin{table}[!t]
\begin{center}
\begin{tabular} {cccc}
\hline
\hline
  Method  &     Est0/True0    &    Est+/True+    &     Est-/True-\\     
   \hline
GSM & 1.009 (2276/2256) & 0.9595 (142/148) & 0.8542 (82/96)\\
Alt-t    & 1.014 (2288/2256) & 0.9189 (136/148) & 0.7917 (76/96)\\
GGM & 1.03 (2324/2256) & 0.8649 (128/148) & 0.5 (48/96)\\
GCGM  & 0.905 (2042/2256) & 1.851 (274/148) & 1.917 (184/96)\\
\hline
\end{tabular}
\end{center}
\caption{Ratio of \#estimated zeros and \#true zeros, \#estimated positive and \#true positive, \#estimated negative and \#true negative elements of the true inverse covariance matrix by the competing methods for continuous non-normal data. Values closer to 1 indicate superior performance. Numbers in parentheses are counts.\label{tab:inv}}
\end{table}

\begin{table}[!t]
\begin{center}
\begin{tabular} {cccc}
\hline
\hline
  Method &     Est0/True0    &    Est+/True+    &     Est-/True-\\     
   \hline
GSM & 1.0346 (2272/2196) & 0.9324 (138/148) & 0.5769 (90/156)\\
GCGM & 0.6029  (1324/2196) & 4.0270 (596/148) & 3.7179 (580/156)\\
\hline
\end{tabular}
\end{center}
\caption{Ratio of \#estimated zeros and \#true zeros, \#estimated positive and \#true positive, \#estimated negative and \#true negative elements of the true inverse covariance matrix by the  proposed method for mixed discrete and continuous data. Values closer to 1 indicate superior performance. Numbers in parentheses are counts.\label{tab:mixed}}
\end{table}

\clearpage\pagebreak\newpage
\thispagestyle{empty}
\begin{center}
{\LARGE{\bf Supplementary Material to\\ {\it Inferring network structure in non-normal and mixed discrete-continuous genomic data}}}
\end{center}

\vskip1cm
\vskip 2mm
\vskip 10mm
\baselineskip=12pt
\begin{center}
Anindya Bhadra\\
Department of Statistics, Purdue University, 250 N. University St., West Lafayette, IN 47907\\
bhadra@purdue.edu\\
\hskip 5mm \\
Arvind Rao\\
Department of Bioinformatics and Computational Biology, The University of Texas MD Anderson Cancer Center, 1400 Pressler Dr., Houston, TX 77030\\
\hskip 5mm \\
Veerabhadran Baladandayuthapani\\
Department of Biostatistics, The University of Texas MD Anderson Cancer Center, 1400 Pressler Dr., Houston, TX 77030  \\
\end{center}

\setcounter{equation}{0}
\setcounter{page}{0}
\setcounter{table}{0}
\setcounter{section}{0}
\setcounter{figure}{0}
\renewcommand{\theequation}{S.\arabic{equation}}
\renewcommand{\thesection}{S.\arabic{section}}
\renewcommand{\thesubsection}{S.\arabic{section}.\arabic{subsection}}
\renewcommand{\thetable}{S.\arabic{table}}
\renewcommand{\thefigure}{S.\arabic{figure}}
\baselineskip=17pt

\clearpage\pagebreak\newpage

\begin{figure}[!t]
\begin{center}
\includegraphics[height=10cm,width=12cm]{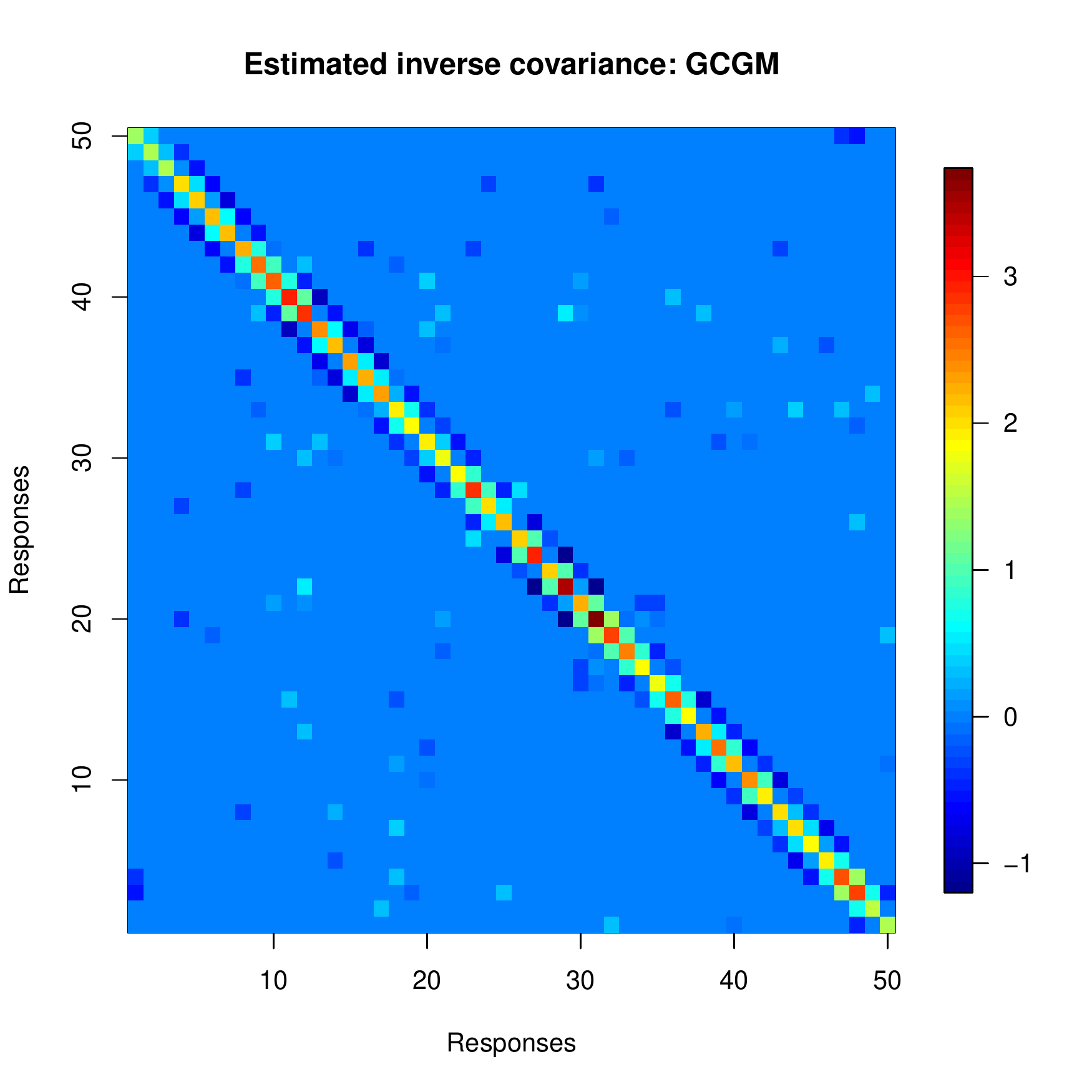}
\end{center}
\caption{Estimated $\Sigma^{-1}$ by Gaussian Copula Graphical Model (GCGM) for continuous non-normal data. \label{fig:supp_heavy}}
\end{figure}

\clearpage\pagebreak\newpage

\begin{figure}[!t]
\begin{center}
\includegraphics[height=10cm,width=12cm]{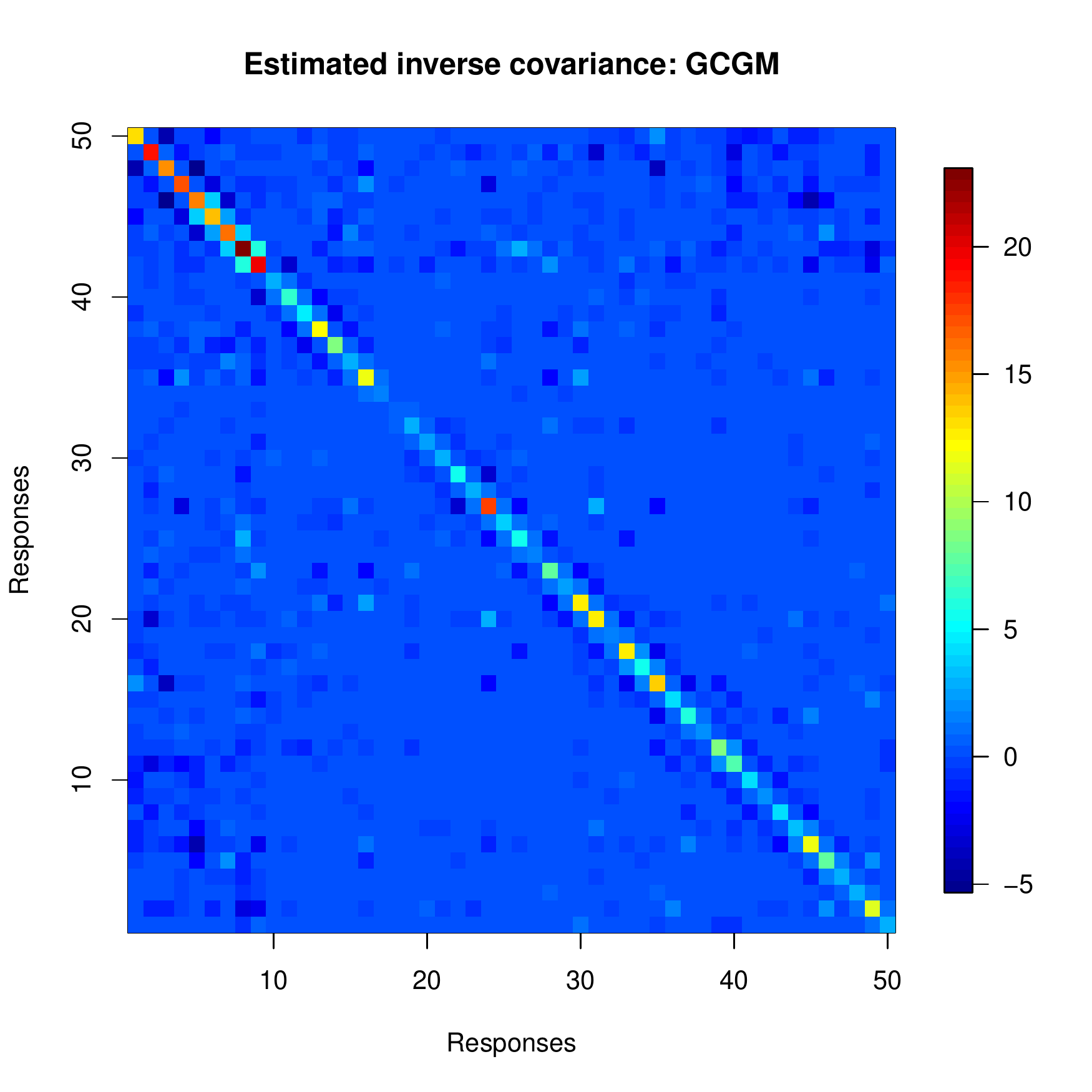}
\end{center}
\caption{Estimated $\Sigma^{-1}$ by Gaussian Copula Graphical Model (GCGM) for mixed discrete and continuous data. \label{fig:supp_mixed}}
\end{figure}

\clearpage\pagebreak\newpage
\begin{table}[!t]
\small
\begin{center}
\begin{tabular} {cc}
\hline
\hline
 Binary Mutations  &   Continuous Expressions\\     
   \hline
'PTEN'
    'TP53'
    'PIK3CA'
    'EGFR'
    'CDKN2C'
    'NF1'
     & 
     'AKT1'
    'AKT2'
    'AKT3'
    'ARAF'
    'BRAF'\\

    'PIK3R1'
    'MDM4'
    'RB1'
    'PIK3C2G'
    &
    'CBL'
    'CCND1'
    'CCND2'
    'CDK4'
    'CDK6'\\
    'MDM2'
    'ERBB2'
    'IRS1'
    'PIK3C2B'
    'CDKN2'
    &
    'CDKN2A'
    'CDKN2B'
    'CDKN2C'
    'EGFR'
    'ERBB2'\\
    
    'PDGFRA'
    'KRAS'
    'PIK3CG'
    'CBL'
    'BRAF'
        &
    'ERBB3'
    'FGFR1'
    'FGFR2'
    'FOXO1A'
    'FOXO3A'\\
    'NRAS'
    'AKT1'
    'PIK3CB'
    'MET'
    'PIK3R2'
       &
    'MLLT7'
    'GAB1'
    'GRB2'
    'HRAS'
    'IGF1R'\\
    'SRC'
    'CCND2'
    'IGF1R'
    'FGFR1'
    
    &
    'IRS1'
    'KRAS'
    'MDM2'
    'MDM4'
    'MET'\\
    &
    'NF1'
    'NRAS'
    'PDGFRA'
    'PDGFRB'
    'PDPK1'\\
    &
    'PIK3C2B'
    'PIK3C2G'
    'PIK3CA'
    'PIK3CB'
    'PIK3CD'\\
    &
    'PIK3CG'
    'PIK3R1'
    'PIK3R2'
    'PTEN'\\
    &
    'RAF1'
    'RB1'
    'SPRY2'
    'SRC'
    'TP53'\\
    \hline
\end{tabular}
\end{center}
\caption{The 29 binary mutations and 49 continuous expression levels considered in the analysis of glioblastoma multiforme data based on 103 patient samples. \label{tab:supp_genes}}
\end{table}

\end{document}